\documentstyle[12pt,epsfig]{article}        
\newlength{\dinwidth}                       
\newlength{\dinmargin}                      
\begin{document}
\begin{flushright}
BRIS/HEP 96-01 \\ 
GLAS-PPE/96-04 \\ 
UCL/HEP 96-03
\end{flushright}
\vspace*{1cm}
\begin{center}  \begin{Large} \begin{bf}
Rapidity Gaps Between Jets\\
  \end{bf}  \end{Large}
  \vspace*{5mm}
  \begin{large}
J. M. Butterworth$^a$, M. E. Hayes$^b$, M. H. Seymour$^c$, L. E. Sinclair$^d$\\ 
  \end{large}
\end{center}
$^a$ University~College~London,~Physics~and~Astronomy~Dept.,~London,~UK \\
$^b$ H.H.~Wills~Physics~Laboratory,~University~of~Bristol,~Bristol,~UK \\
$^c$ TH Division, CERN, CH-1211 Gen\`eve 23, Switzerland \\
$^d$ Dept.~of~Physics~and~Astronomy,~University~of~Glasgow,~Glasgow,~UK \\
\begin{quotation}
\noindent
{\bf Abstract:}
An excess of events with a rapidity gap between jets, over what would
be expected from non-diffractive processes, has been observed at HERA.
A process based on a perturbative QCD calculation of colour singlet
exchange has been added to HERWIG. With this addition, HERWIG is 
able to describe the number of events with a gap between jets over the 
number without a gap. This gap fraction is predicted to rise at large 
rapidity intervals between jets which would only be visible if the 
detector coverage were increased.
\end{quotation}
\begin{figure}[htb] \label{FIG1}
\centering
\epsfig{file=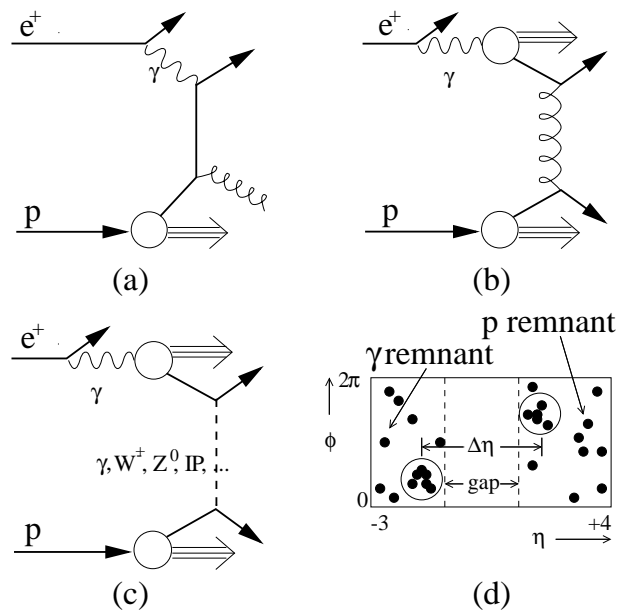,height=8.0cm}
\caption[junk]{{\it
Feynman diagrams of processes involved. (a) and (b) are examples of the
direct and resolved contributions in photoproduction. (c) is the colour 
singlet exchange. (d) is the representation of how an individual event would
appear in $\eta$--$\phi$ space.
  }}
\end{figure}

At HERA the incoming electron is accompanied by a cloud of photons.
The spectrum of photon virtuality, $P^2$, is dominated by almost-real
photons and, provided $P^2$ is prevented from being too large, the
incoming photon can be regarded as real and the types of event it
participates in are classified as photoproduction events.  We use
the usual experimental cut of $P^2<4GeV^2$.
If the event contains one or more jets it is assumed to have a hard
interaction and is then perturbatively calculable in QCD; hence it is
referred to as a hard photoproduction event. 
Here we will refer only to these kinds of events. These 
can be of two main types: direct and resolved photoproduction,
shown in figures 1(a) and (b). In the direct case the whole photon interacts 
with a parton in the proton.
In the resolved case the 
photon `resolves' into partons, one of which then interacts with a 
parton from the proton.  
If the transverse momentum exchanged is high enough, outgoing partons
give rise to `jets' of particles in the detector.

 In most cases of parton-parton scattering a coloured object is exchanged
 (e.g. fig.~1(b)) with an associated colour flow between the outgoing partons
 and the remnant particles.  This can be modelled by a colour string
 which is stretched across the central rapidity interval.  This string
 then fragments into particles which occupy the region between the two
 jets.  However it is also possible for the exchanged object to be a colour
 singlet (fig.~1(c)).  In this case the colour strings connect each outgoing
 parton with the remnant jet closest to it in rapidity.  This leads to a
 suppression of particle production in the rapidity region between the two
 jets.  It was first observed at $p\bar p$ colliders \cite{D0} 
 and subsequently measured at HERA \cite{Sinclair}.
 
The presence of high $E_T$ jets in these events guarantees a high value
 of $\hat{t}$ and assures us of the applicability of perturbative QCD,
$-\hat{t}/\Lambda^2_{QCD} \gg 1$
($\hat{s}$ is the centre of mass energy and
$\hat{t}$ the invariant momentum transfer squared of the parton system). 
In addition, the pseudorapidity interval between the
jets, $\Delta\eta$, reflects the separation in rapidity of the outgoing
partons, $y$.  Therefore at large $\Delta\eta$ we have
$y \simeq \ln (- \hat{s} / \hat{t})$ and $-\hat{s}/\hat{t} \gg 1$.
Gluon exchange in the $\hat{t}$-channel increases with $1/\hat{t}^2$. 
However because a gluon is a coloured object, a rapidity gap does not
normally result. Two gluon exchange, which may be
in a colour singlet state, can give rise to a rapidity gap.
Gluons exchanged between this $\hat{t}$--channel gluon pair can further
enhance the cross section such that it rises faster than $1/\hat{t}^2$ at
small $\hat{t}$.  These gluons can be summed using the BFKL equation 
\cite{BFKL}, to give
$$ {\mathrm{d}\sigma ( qq \rightarrow qq) \over \mathrm{d}\hat{t}} = 
( \alpha_s C_F )^4 {\pi^3 \over 4 \hat{t}^2} 
{\exp (2 \omega_0 y) \over [ {7\over 2} \alpha_s C_A \zeta (3) y]^3}, $$
as derived in ref.~\cite{Mueller}
($C_F = 4/3$, $C_A = N_c =3$, 
$\zeta (3) \simeq 1.202$ is the Riemann zeta function and 
$\omega_0 = {C_A \alpha_s \over \pi} 4 \ln 2$).
The approximations made in deriving this formula mean that the correct
scale to use in $\alpha_s$ cannot be determined, and does not even need
to be the same in each case.  We use $\alpha_s(-\hat{t})$ in the
prefactor, $\alpha_s=0.25$ in the denominator and $\omega_0=0.3$
(which have been installed as HERWIG defaults).


The package HzTool \cite{HzTool} was used to generate the Monte Carlo and to 
compare to the original analysis \cite{Sinclair} on ZEUS 1994 data. 
This facilitated 
the comparison of the data to the Monte Carlo generated. HERWIG 5.8d 
\cite{HERWIG} was upgraded to include the QCD calculation \cite{Mueller} 
described above. 
About $2.6\mathrm{pb}^{-1}$ (the 1994 ZEUS
luminosity) of events were then generated.  A cone-based jet finding
algorithm was run on this sample.  Events were required to have at least
two jets of $E_T^{jet} > 6$ GeV
with pseudorapidity satisfying $\eta_{jet} < 2.5$.
Denoting by $\eta_1$ and $\eta_2$ the pseudorapidities of the two highest
$E_T$ jets, the events were required to satisfy in addition
$|{\eta_1 + \eta_2\over 2}| < 0.75$ and $\Delta\eta = \eta_1 - \eta_2 > 2$.
Events with no particle with transverse energy greater than
300 MeV between the two jets were then classified as `gap' events.
The characteristic signature of these events is illustrated in fig.~1(d).

The selected sample exhibits an exponential decay in transverse momentum
(fig.~2(a)) and a bias towards high $x_\gamma^{OBS}$ (fig.~2(b)).
$x_\gamma^{OBS}$ is the 
fraction of momentum of the photon that participates in the interaction as 
calculated from
the observed jets. The differential cross section as a function of $\Delta\eta$
is also shown (fig.~2(c)). The gap fraction (fig.~2(d)) is the number of events 
with a gap divided by the total number of dijet events. It levels off at
around $60\%$. Although a fraction of $100\%$ might be expected, many gap events
are lost by such factors as the final state particles escaping the bounds 
defined by the jet cone and hence filling the gap.
\begin{figure}[htb] \label{FIG2}
\centering
\epsfig{file=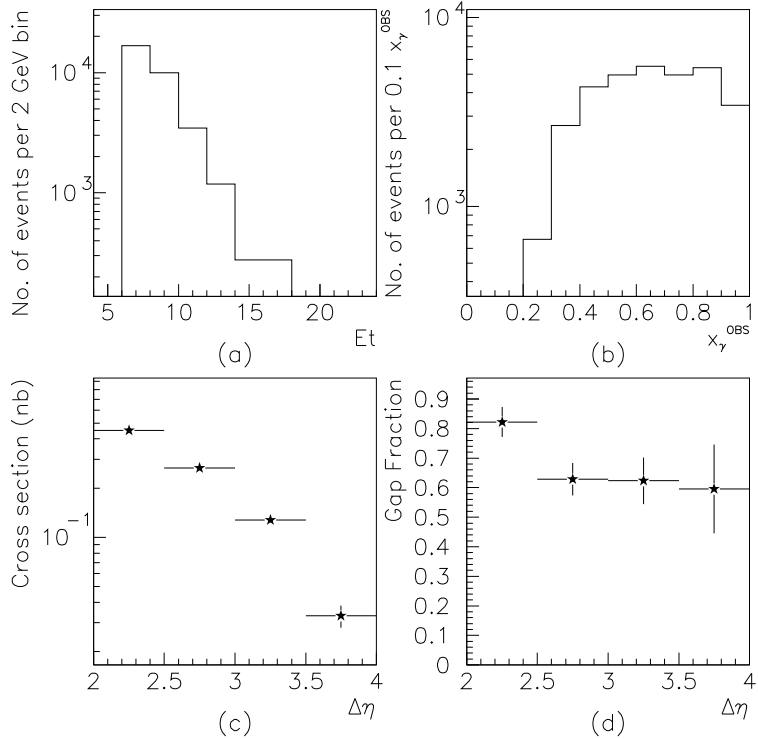,height=10cm}
\caption[junk]{{\it
The QCD colour singlet process by itself. (a) shows the spectrum of the jet
$E_T$
used in the sample. (b) shows the $x_\gamma^{OBS}$ distribution of
the sample. (c) shows the cross section of the sample as producd by HERWIG,
in bins of $\Delta\eta$ (the interval between jets).
(d) shows the fraction of events with a gap over the total number of dijet
events.
  }}
\end{figure}

HERWIG 5.8d was then used to produce `standard' direct and resolved
photoproduction events and these were added to the sample of `colour singlet'
events. 
The cross section for the dijet sample without the requirement for a jet
was tuned to the ZEUS 1994 data.
This was then held fixed as the
`colour singlet' sample normalization was adjusted to fit the 
gap fraction graph. 
An overall normalization factor of 30 for the `colour singlet' sample
was found necessary to describe the data. This factor is allowed due to
theoretical uncertainties in the value of $\alpha_s$ as mentioned above.
PYTHIA 5.7 \cite{PYTHIA}
was then used to produce `standard' events, as a comparison.

Both Monte Carlos are compared to the
data (fig.~3(a)). As can be seen from the last two bins, PYTHIA, which does
not include colour singlet exchange fails to
describe the data. However HERWIG, with the new QCD process, provides a good
description. Also included is the standard HERWIG plot (HERWIG 5.8d) without
the colour singlet sample. This emphasizes the dramatic effect, at high 
$\Delta\eta$ of the new process on the gap fraction.
\begin{figure}[htb] \label{FIG3}
\centering
\epsfig{file=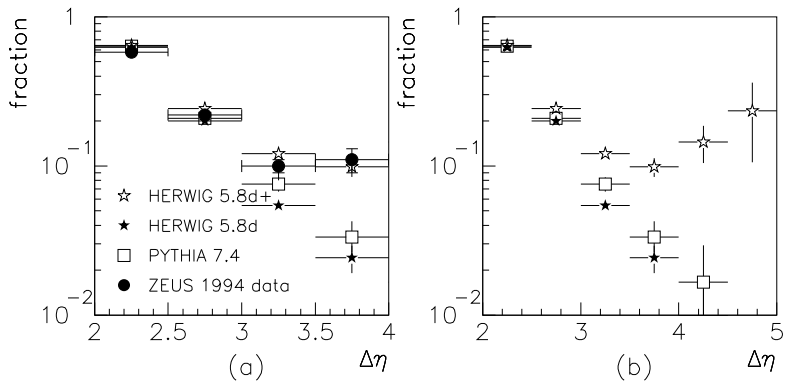,height=5cm}
\caption[junk]{{\it
Gap fractions as produced by PYTHIA and HERWIG (with the colour singlet
interaction denoted by 5.8d+). (a) is for a standard
detector, and the ZEUS 1994 data has been superimposed. (b) is for
an extended coverage detector.
  }}
\end{figure}

Using this model the Monte Carlo was used to simulate an extended detector.
The extended detector coverage was 
simulated by changing the $\eta_{jet}$ cut to $\eta_{jet} < 3.5$. 
With the extended detector
clear evidence for the colour singlet exchange for $\Delta\eta>4.0$
 is seen (fig.~3(b)).
Figure~3 also shows a characteristic rise in the gap fraction. The
gap fraction of the process by itself is~$60\%$. The diffractive cross section
falls less rapidly as $\Delta\eta$ increases than does the `background'
non-diffractive cross section. Thus at large $\Delta\eta$ we expect the
gap fraction to rise towards $60\%$. Hence given an
increased detector acceptance, we can see a clear signal for this kind of
event.

The extended HERWIG was also used to simulate increased luminosity with
the current detector coverage. Although increased statistics would allow
different bins of $E_T$ of the jets and of $x_\gamma^{OBS}$ no such striking
signal of the colour singlet exchange would be seen. So we conclude in
favour of the extended detector coverage.

Thanks to Jeff Forshaw for useful discussions.

\end{document}